\documentclass[12pt,a4paper]{article}
\usepackage{graphics}
\usepackage{amsmath}
\numberwithin{equation}{section}
\title{Verhulst-Lotka-Volterra (VLV) model of ideological struggles}
\author{Marcel R. Ausloos$^{1,}$\footnote{e-mail: marcel.ausloos@ulg.ac.be} Nikolay K. Vitanov$^{2,}$\footnote{e-mail: vitanov@imbm.bas.bg}, Zlatinka I. Dimitrova$^{3}$,}   
\date{ \small $^{1}$ 7 rue des Chartreux, B-4122 Neupr\'e, Wallonia,\\ {\it previously at} GRAPES, Institute of Physics B5a, University of Li\`ege, \\ B-4000 Li\`ege,
Euroland
\\$^{2}$Institute of Mechanics, Bulgarian Academy of Sciences,\\
Akad. G. Bonchev Str., Bl. 4, \ 1113 Sofia, Bulgaria, %\footnote{e-mail: vitanov@imbm.bas.bg}
 \\
$^{3}$Institute of Solid State Physics, Bulgarian Academy of Sciences, \\
Blvd. Tzarigradsko Chaussee 72, 1784, Sofia, Bulgaria \\
}

\begin{document}
\maketitle
\begin{abstract}
Let the   population of  e.g. a country where  some opinion struggle occurs
  be varying in time, according to Verhulst equation.  Consider next some  competition between opinions  such as the dynamics  be described by  Lotka and Volterra equations.  
Two kinds of influences can be used, in such a model,  for  describing the dynamics of an agent opinion conversion:  this can occur (i)   either by means of mass 
communication tools,  under some external field influence,   or (ii)  by means of direct interactions between agents.
It results, among other features, that  change(s) in environmental conditions can prevent the
extinction of populations of followers of some ideology due to different kinds of   resurrection effects.
The tension arising in the  country population  is proposed to be measured by an appropriately defined scale index. \end{abstract}
\begin{flushleft}

%PACS: 87.23.Ge - Dynamics of social systems; 87.23.Cc Population dynamics and ecological pattern formation
\end{flushleft}
\section{ Introduction %Opinion formation and population dynamics
}
\label{intro}

 Not  further elaborating here  on riots,  revolutions,  or wars, it is well known that socio-economic systems, or civilisations,  survive because of the resolution of conflicts.  These are indeed sources of population decimation, and of various critical events, called "humanitarian crises".
 
 Opinion conflicts, so numerous in economic and social systems, are thus  connected to $competition$ between so called degrees of freedom:
religions \cite{petr07e,petr07a}, languages \cite{abrams}, businesses   \cite{caram}, and  to the problem of $growth/extinction$ of populations \cite{pec08}, {\it per se}. 
Whence the  need for a  basic  theoretical approach
and some modelling in order to focus any data gathering toward useful input in further work.
 
 In this spirit, an analytical  model for ideological struggles in e.g. economy and social systems is here below 
formulated. The underlying set of agents is supposed to be a closed and finite one, like those usually existing in a so called country, but we emphasise letting the population 
size be varying in time for being more realistic. An appropriate equation for doing so seems that of Verhulst \cite{Verhulst}, extending the classical equation of exponential growth \cite{Euler,Malthus}. The dynamics of the struggle is next to be modelled; here below it is  described  through model
equations   as those of Lotka and Volterra \cite{Lotka,Volterra}, Thus the analytical  model of competition between opinions is 
formulated through a combined set of well known equations, each set based on one variable, i.e. for each ideology population (or rather concentration)\footnote{During the writing of this paper, we have noticed a simplified version of the present model; see \cite{ZwanzigPNAS73}}.

 For being imaginative, notice that
models other than that of Volterra and Lotka can be used as
starting points. For example, Montroll \cite{M6} has suggested a
model in which the logarithms of populations obey linear
equations. This model seems to be connected with Gompertz's
law of population growth \cite{Gompertzlaw} in the same way that the
 Volterra-Lotka model is connected with Verhulst's (so called logistic) law.  
 Moreover discrete simulation approaches can be envisaged: see an interesting but brief review on  prey-predator  models on lattices   by Pekalski in \cite{pekComSim2004}

Most importantly: several  ''ideologies''  usually compete to increase their number of adepts.
Members or "followers"  can be either converted from one ideology to another or become followers of one 
ideology though being previously ideologically-free. Reverse processes can be also allowed, with a so called change of mind.

Two kinds of influences can be used for conversion: (i)   conversion by means of mass 
communication tools, thus under some external field influence, or  (ii) conversion  by means of direct interactions between agents. These different processes  should remind one of the various terms in Hamiltonian mechanics based on the level  of interactions between entities., i.e. the number of clusters to be considered, as in a renormalisation group approach. Moreover those terms are the basic ones in a Lotka-Volterra equation, one is linear in the population concentration, the other based on the product of the population concentrations.

As interesting results,  it is briefly mentioned that  change(s) in environmental conditions can prevent the
extinction of populations of followers of some ideology. Instead  different kinds of   resurrection effects can occur, so called phoenix effects.
Moreover when a new ideology appears in the system, since some tension arises, stressing the  population steady state(s), it is proposed to measure the ideology tensions through an appropriately defined scale index. The latter evolution would be a convenient  monitoring measure for calibrating the model parameters.

\section{Mathematical formulation of the model}
\label{sec:2}
Consider    $N$ agents, divided into $n+1$  different specific ideologies,  such that  the
number of members in the corresponding populations  are $N_{1}, N_{2}, \dots, N_{n}$, and   also 
 $N_{0}$  agents who are not followers of  any ideology at a given 
  time $t$, or i.e. $
N =  \sum_{i=0}^{n} N_{i}.
$
Moreover define
 $C(t,N,N_{\nu},p_{\mu})$ as
 the   so called carrying capacity; let  $p_{\mu}$ stand  for $(p_{1},\dots, p_{m})$    parameters describing the environment  and   
$N_{\nu} $ for $ (N_{0}, N_{2}, \dots, N_n)$.

The overall population is supposed to
evolve according to% the generalized Verhulst law
\begin{eqnarray}\label{N}
\frac{dN(t)}{dt}=r(t,N,N_{\nu},p_{\mu}) N(t) \left[ 1 - \frac{N(t)}{C(t,N,N_{\nu},p_{\mu})} \right]
\end{eqnarray}
where the growth process is   
monitored   \cite{wilke} through the overall population growth  rate
%\footnote{Optimistically, let us only consider the case $r>0$.}  
$r(t, N, N_{\lambda}, 
p_{\mu}) \ge0$.

For every ideology/population $i$  one should  account for  
(1) deaths, (2) unitary conversion, and (3) binary conversion:

\begin{enumerate}

\item
  Death implies a term $r_{i} N_{i}$, with $r_{i}  =r_{i}(t,N,N_{\nu},p_{\mu}, C)\le 0$.
\item
Unitary conversion  
is made  only 
  through the information environment of the population. The number of people converted from ideology $j$ to ideology $i$
is  assumed to  be proportional to the number $N_{j}$.  An 
$f_{ij}=f_{ij}(t,N,N_{\nu},p_{\mu},C)$ coefficient characterises the intensity  with which this conversion occurs; $f_{ii}=0$.
The corresponding modelling term is $f_{ij} N_{j}$

\item
Binary conversion occurs when there is conversion to the
$i$-th ideology because of direct interaction 
between members of the $j$-th and $k$-th ideology. 
It is assumed that the intensity of the interpersonal contacts is
proportional to the numbers $N_{j}$ and $N_{k}$. 
The coefficient characterising the intensity of the binary conversion  is  $b_{ijk}=b_{ijk}(t,N,N_{\nu},p_{\mu},C)$;
of course $b_{iii}=0$.
The corresponding modelling term is  $ b_{ijk} N_{j} N_{k}$.

\end{enumerate}

For space saving, the final set of equations is not further written, but the case of
constant values of the parameters is discussed in the next section. 
Notice that
arbitrary values are not allowed for the coefficients of the model; e.g. they must 
have   values  such that  $N$, $N_{0}$,  $\dots$, $N_{n}$   be positive. 
 
\section{Summary of Results}

Analytical and simulation studies have been performed for constant values of the parameters,  i.e.  the struggling population $i$ evolution is governed by
\begin{equation}\label{mod_syst3}
\frac{dN_{i}}{dt} = r_{i} N_{i}  + \sum_{j=0}^{n} f_{ij} N_{j}+
 \sum_{j=0}^{n} \sum_{k=0}^{n} b_{ijk} N_{j} N_{k} \;,
\end{equation}
bearing Eq.(\ref{N}) as a constraint $on$ the population competition \cite{dv01a,dv06}. Notice also that terms like $f_{i0} N_{i0}$, or  $b_{ii0} N_{i} N_{0} $ or $b_{i00} N_{0}^2 $, etc. are not excluded
\begin{itemize}

\item  
For the case of one ideology, the  %discussed 
simple version of the general model
describes the evolution to an equilibrium state in which the population 
consists of some amount of  followers of the ideology and
persons indifferent to the ideology.
 The case of one ideology in presence of a background population reaches an asymptotic stable state.  An
 interesting situation occurs when $\frac{d N_{1}}{dt}$   changes  sign with
  $t$. It can be observed that the number of followers of the
ideology can increase despite the fact that $r_{1}<0$. The reason for this effect
is the $N_{0}$ increase, occurring  because of the fast
 growth of the country population. When $N_{0}$ is small the term containing
$r_{1}$ dominates and $N_{1}$ decreases. But in the course of time $N_{0}$ increases.
Then the conversion begins to dominate over dissatisfaction and the number of  
followers of the ideology   increases.
Such a  growth is   called   an $ inertial$ growth.

 \item When two ideologies compete in presence of the $ideology-free$ population, it can  be observed  that  an ideology population  shrinks.
At the fixed point, e.g. 
\begin{equation}\label{fixpoints1}
\hat{N}_{1} = \frac{C f_{10}}{f_{10}-r_{1}} \rightarrow
\breve{N}_{1}=\frac{C r_{2} f_{10}}{r_{2}(f_{10} - r_{1})+r_{1} f_{20} },
\end{equation}

Moreover, it can be   observed  that  change(s) in environmental conditions, through the set of parameteres $p_{\mu}$, explicitly or implicitly appearing in $C$,  $r_{i} $, $ f_{ij} $, 	and $ b_{ijk} $, can prevent the
extinction of populations of followers of some ideology. Different kinds of   resurrection (so called phoenix) effects can occur, depending on the most relevant term, in particular  $f_{ij}$ or $b_{ijk}$.

It is suggested that a measure of the tension between populations, with different opinions,  can be through the index
 \begin{equation}\label{strain_meas1}
T_{i;k}(t) = 1 - \frac{N_{i}^{(k)}(t)}{\hat{N}_{i}},
\end{equation}
where $N_{i}^{(k)}(t)$ is the population of the followers of the $i$-th ideology when
the $k$-th ideology is present.

 \item  The case of  three  and four  ideologies has been numerically investigated elsewhere \cite{VDA10},  also
 generalising the notion of tension index  for the case of an  arbitrary number of ideologies
in the country \cite{VDA10}.   Steady states, cycles or   chaotic regimes can be obtained.

\end{itemize}

\section{Concluding remarks}
 
Notice that the model applies to cases in which countries have a  variable total population
which evolves according to the generalised Verhulst model.   Moreover it should be  emphasised that a conversion can be outside the  
competing ideologies of interacting agents: see Eq.(\ref{mod_syst3}) last term.
 
In contrast to simulation model, avoiding stochastic noise as an input,  the above analytical model allows for resurrection ("phoenix") effects. The introduction of an ideology
leads to some tension between the previously present ideologies, as the numbers of  followers drop. Such tensions can be quantified by a set of indices. 
Although the number of parameters seems considerable, as in many realistic population 
evolution studies,  the set of parameters appears to be realistic enough to be 
calibrated  in specific situations, in particular by monitoring  such tension indices.
 
\vskip 0.5truecm

 {\bf Acknowledgements}  \\
 Thanks to the  COST MP0801 "Physics of Competition and Conflict" Action
for support of our research. 

 % Non-BibTeX users please use

\end{document}